\documentclass[aps,prl,twocolumn,showpacs,superscriptaddress]{revtex4}

\usepackage{graphicx} 

\begin{document}

\title
{Half-metallic silicon nanowires}
\author{E. Durgun}
\affiliation{Department of Physics, Bilkent University, Ankara
06800, Turkey}
\affiliation{National Nanotechnology Research
Center, Bilkent University, Ankara 06800, Turkey}
\author{D. \c{C}ak{\i}r}
\affiliation{Department of Physics, Bilkent University, Ankara
06800, Turkey} \affiliation{National Nanotechnology Research
Center, Bilkent University, Ankara 06800, Turkey}
\author{N. Akman}
\affiliation{National Nanotechnology Research Center, Bilkent
University, Ankara 06800, Turkey} \affiliation{Department of
Physics, Mersin University, Mersin, Turkey}
\author{S. Ciraci} \email{ciraci@fen.bilkent.edu.tr}
\affiliation{Department of Physics, Bilkent University, Ankara
06800, Turkey}
\affiliation{National Nanotechnology Research
Center, Bilkent University, Ankara 06800, Turkey}

\date{\today}

\begin{abstract}
From first-principles calculations, we predict that transition
metal (TM) atom doped silicon nanowires have a half-metallic
ground state. They are insulators for one spin-direction, but show
metallic properties for the opposite spin direction. At high
coverage of TM atoms, ferromagnetic silicon nanowires become
metallic for both spin-directions with high magnetic moment and
may have also significant spin-polarization at the Fermi level.
The spin-dependent electronic properties can be engineered by
changing the type of dopant TM atoms, as well as the diameter of
the nanowire. Present results are not only of scientific interest,
but can also initiate new research on spintronic applications of
silicon nanowires.
\end{abstract}



\maketitle

Rod-like, oxidation resistant Si nanowires (SiNW) can now be
fabricated at small diameters\cite{ma} (1-7 nm) and display
diversity of interesting electronic properties. In particular,
the band gap of semiconductor SiNWs varies with their diameters.
They can serve as a building material in many of electronic and
optical applications like field effect transistors \cite{Cui}
(FETs), light emitting diodes \cite{Huang}, lasers \cite{Duan} and
interconnects. Unlike carbon nanotubes, the conductance of
semiconductor nanowire can be tuned easily by doping during the
fabrication process or by applying a gate voltage in a SiNW FET.

In this letter, we report a novel spin-dependent electronic
property of hydrogen terminated silicon nanowires (H-SiNW): When doped
by specific transition metal (TM) atoms they show half-metallic\cite{groot,pickett}
(HM) ground state. Namely, due to broken spin-degeneracy, energy bands
$E_{n}({\bf k },\uparrow)$ and $E_{n}({\bf k},\downarrow)$ split
and the nanowire remains to be insulator for one spin-direction
of electrons, but becomes a conductor for the opposite spin-direction
achieving 100\% spin polarization at the Fermi level. Under certain
circumstances, depending on the dopant and diameter,
semiconductor H-SiNWs can be also either a ferromagnetic semiconductor
or metal for both spin directions. High-spin polarization at the Fermi
level can be achieved also for high TM coverage of specific SiNWs.
Present results on the asymmetry of electronic states of TM doped SiNWs
are remarkable and of technological interest since room temperature
ferromagnetism is already discovered in Mn-doped SiNW\cite{wu}.
Once combined with advanced silicon technology, these properties can
be realizable and hence can make "known silicon" again a potential
material with promising nanoscale technological applications in
spintronics, magnetism.

Even though 3D ferromagnetic Heusler alloys and transition-metal oxides exhibit
half-metallic properties \cite{park}, they are not yet
appropriate for spintronics because of difficulties in controlling
stoichiometry and the defect levels destroying the coherent
spin-transport. Qian \emph{et al.} have proposed HM
heterostructures composed of $\delta$-doped Mn layers in bulk Si
\cite{fong}. Recently, Son \emph{et al.} \cite{cohen} predicted HM
properties of graphene nanoribbons. Stable 1D half-metals have
been also predicted for TM atom doped arm-chair single-wall carbon
nanotubes \cite{yang} and linear carbon chains \cite{sefa,engin};
but synthesis of these nanostructures appears to be difficult.

Our results are obtained from first-principles plane wave
calculations \cite{vasp} (using a plane-wave basis set up to
kinetic energy of 350 eV) within generalized gradient
approximation expressed by PW91 functional\cite{gga}. All
calculations for paramagnetic, ferromagnetic and antiferromagnetic
states are carried out using ultra-soft pseudopotentials
\cite{vanderbilt} and confirmed by using PAW
potential\cite{blochl}. All atomic positions and lattice constants
are optimized by using the conjugate gradient method where total
energy and atomic forces are minimized. The convergence for energy
is chosen as 10$^{-5}$ eV between two steps, and the maximum force
allowed on each atom is 0.05 eV/\AA \cite{detail}.

\begin{figure}
\includegraphics[scale=0.45]{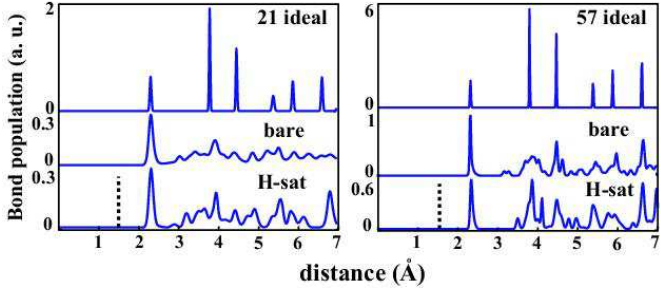}
\caption{(Color online) Upper curve in each panel with numerals
indicate the distribution of first, second, third, fourth etc
nearest neighbor distances of SiNW(N) as cut from the ideal Si
crystal, same for structure-optimized bare SiNW(N)(middle curve)
and structure optimized H-SiNW(N) (bottom curve) for N=21, 57 and
81. Vertical dashed line corresponds to the distance of Si-H
bond.} \label{fig:bond1}
\end{figure}

\begin{figure}
\includegraphics[scale=0.55]{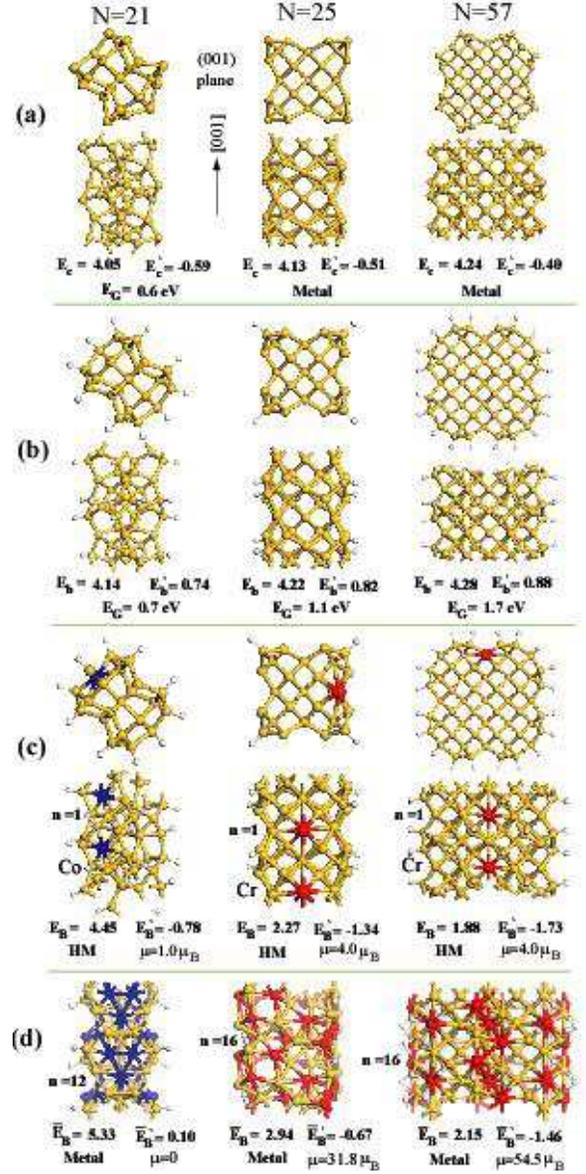}
\caption{(Color online) Top and side views of optimized atomic
structures of various SiNW(N)'s. (a) Bare SiNWs; (b) H-SiNWs;
(c) single TM atom doped per primitive cell of H-SiNW ($n=1$); (d)
H-SiNWs covered by $n$ TM atom corresponding to
$n>1$. $\overline{E}_{c}$, $\overline{E}^{\prime}_{c}$, $E_{b}$,
$E^{\prime}_{b}$, $E_{G}$, and $\mu$, respectively denote the
average cohesive energy relative to free Si atom, same relative to the
bulk Si, binding energy of hydrogen atom relative to free H atom,
same relative to H$_{2}$ molecule, energy band gap and the net
magnetic moment per primitive unit cell. Binding energies in
regard to the adsorption of TM atoms, i.e. $E_{B}$,
$E^{\prime}_{B}$ for $n=1$  and average values
$\overline{E}_{B}$, $\overline{E}^{\prime}_{B}$ for $n >1$
are defined in the text and in Ref[\onlinecite{binding}]. The [001]
direction is along the axis of SiNWs. Small, large-light and
large-dark balls represent H, Si and TM atoms, respectively. Side
views of atomic structure comprise two primitive unit cells of the
SiNWs. Binding and cohesive energies are given in eV/atom.}
\label{fig:structure}
\end{figure}

Bare SiNW(N)s (which are oriented along [001] direction and have N
Si atoms in their primitive unit cell) are initially cut from the
ideal bulk Si crystal in rod-like forms and subsequently their
atomic structures and lattice parameter are relaxed \cite{sinw}.
The optimized atomic structures are shown for N=21, 25, and 57 in
Fig. \ref{fig:structure}. While bare SiNW(21) is a semiconductor,
bare SiNW(25) and SiNW(57) are metallic. The average cohesive
energy relative to a free Si atom ($\overline{E}_{c}$) is
comparable with the calculated cohesive energy of bulk crystal
(4.64 eV per Si atom) and it increases with increasing N. The
average cohesive energy relative to the bulk Si crystal,
$\overline{E}^{\prime}_{c}$, is small but negative as expected.
Upon passivation of dangling bonds with hydrogen atoms all of
these SiNWs (specified as H-SiNW) become semiconductor with a band
gap $E_{G}$. The binding energy of adsorbed hydrogen relative to
the free H atom ($E_{b}$), as well as relative to the free H$_{2}$
($E^{\prime}_{b}$) are both positive and increases with increasing
N. Extensive \emph{ab initio} molecular dynamics calculations have
been carried out at 500 K using supercells, which comprise either
two or four primitive unit cells of nanowires to lift artificial
limitations imposed by periodic boundary condition. After several
iterations lasting 1 ps, the structure of all SiNW(N) and
H-SiNW(N) remained stable. Even though SiNWs are cut from ideal
crystal, their optimized structures deviate substantially from
crystalline coordination, especially for small diameters as seen
in Fig.\ref{fig:bond1}. Upon hydrogen termination the structure is
healed substantially, and approaches the ideal case with
increasing N (or increasing diameter), as expected. The calculated
response of the wire to a uniaxial tensile force, $\kappa=\partial
E_{T}/\partial c$, ranging from 172 to 394 eV/cell indicates that
the strength of H-SiNW(N)s (N=21-57) is rather high.

The adsorption of a single TM (TM=Fe, Ti, Co, Cr, and Mn) atom per
primitive cell, denoted by $n=1$, have been examined for different
sites (hollow, top, bridge etc) on the surface of H-SiNW(N) for
N=21, 25 and 57. In Fig. \ref{fig:structure}(c) we present only
the most energetic adsorption geometry for a specific TM atom for
each N, which results in a HM state. These are Co-doped
H-SiNW(21), Cr-doped SiNW(25) and Cr-doped SiNW(57). These
nanowires have ferromagnetic ground state, since their energy
difference between calculated spin-unpolarized and spin-polarized
total energy, i.e. $\Delta E^{m} = E^{su}_{T} - E^{sp}_{T}$ is
positive. We calculated $\Delta E^{m}=$0.04, 0.92 and 0.94 eV for
H-SiNW(21)+Co, H-SiNW(25)+Cr and H-SiNW(57)+Cr, respectively
\cite{magnet}. Moreover, these wires have the integer number of
unpaired spin in their primitive unit cell. In contrast to usually
weak binding of TM atoms on single-wall carbon nanotubes which can
lead to clustering \cite{engin2}, the binding energy of TM atoms
($E_{B}$) on H-SiNWs is high and involve significant charge
transfer from TM atom to the wire \cite{binding}. Mulliken
analysis indicates that the charge transfer from Co to H-SiNW(21)
is 0.5 electrons. The charge transfer from Cr to H-SiNW(25) and
H-SiNW(57) is even higher (0.8 and 0.9 electrons, respectively).
Binding energies of adsorbed TM atoms relative to their bulk
crystals ($E^{\prime}_{B}$) are negative and hence indicate
endothermic reaction. Due to very low vapor pressure of many
metals, it is probably better to use some metal-precursor to
synthesize the structures predicted here.

\begin{figure}
\includegraphics[scale=0.65]{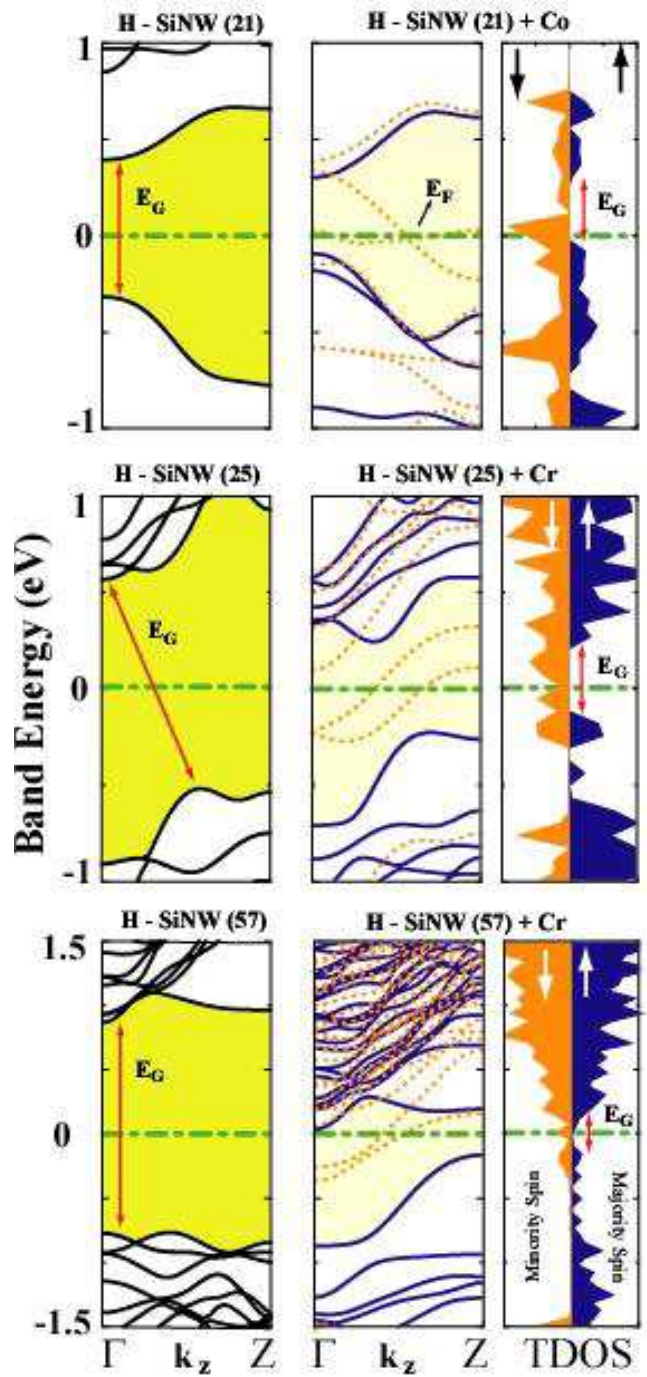}
\caption{(Color online) Band structure and spin-dependent total
density of states (TDOS) for N=21, 25 and 57. Left panels:
Semiconducting H-SiNW(N). Middle panels: Half-metallic
H-SiNW(N)+TM. Right panels: Density of majority and minority spin
states of H-SiNW(N)+TM. Bands described by continuous and dotted
lines are majority and minority bands. Zero of energy is set to
$E_{F}$.} \label{fig:bands}
\end{figure}

The band structures of HM nanowires are presented in
Fig.\ref{fig:bands}. Once a Co atom is adsorbed above the center
of a hexagon of Si atoms on the surface of H-SiNW(21) the spin
degeneracy is split and whole system becomes magnetic with a
magnetic moment of $\mu$=1 $\mu_{B}$ (Bohr magneton per primitive
unit cell). Electronic energy bands become asymmetric for
different spins: Bands of majority spins continue to be
semiconducting with relatively smaller direct gap of $E_{G}$=0.4
eV. In contrast, two bands of minority spins, which cross the
Fermi level, become metallic. These metallic bands are composed of
Co-$3d$ and Si-$3p$ hybridized states with higher Co contribution.
The density of majority and minority spin states, namely
$D(E,\uparrow)$ and $D(E,\downarrow)$, display a 100\%
spin-polarization
$P=[D(E_{F},\uparrow)-D(E_{F},\downarrow)]/[D(E_{F},\uparrow)+D(E_{F},\downarrow)]$
at $E_{F}$. Cr-doped H-SiNW(25) is also HM. Indirect gap of
majority spin bands has reduced to 0.5 eV. On the other hand, two
bands constructed from Cr-$3d$ and Si-$3p$ hybridized states cross
the Fermi level and hence attribute metallicity to the minority
spin bands. Similarly, Cr-doped H-SiNW(57) is also HM. The large
direct band gap of undoped H-SiNW(57) is modified to be indirect
and is reduced to 0.9 eV for majority spin bands. The minimum of
the unoccupied conduction band occurs above but close to the Fermi
level. Two bands formed by Cr-$3d$ and Si-$3p$ hybridized states
cross the Fermi level. The net magnetic moment is 4 $\mu_{B}$.
Using PAW potential results, we estimated Curie temperature of
half-metallic H-SiNW+TMs as 8, 287, and 709 K for N=21, 25, and
57, respectively.

The well-known fact that density functional theory underestimates
the band gap, $E_{G}$ does not concern the present HM states,
since H-SiNWs are already verified to be semiconductor
experimentally\cite{ma} and upon TM-doping they are predicted to
remain semiconductor for one spin direction. In fact, band gaps
predicted here are in fair agreement with experiment and theory.
As for partially filled metallic bands of the opposite spin, they
are properly represented. Under uniaxial compressive strain the
minimum of the conduction band of majority spin states rises above
the Fermi level. Conversely, it becomes semi-metallic under
uniaxial tensile strain. Since conduction and valence bands of
both H-SiNW(21)+Co and H-SiNW(25)+Cr are away from $E_{F}$, their
HM behavior is robust under uniaxial strain. Also the effect of
spin-orbit coupling is very small and cannot destroy HM properties
\cite{sefa}. The form of two metallic bands crossing the Fermi
level eliminates the possibility of Peierls distortion. On the
other hand, HM ground state of SiNWs is not common to all TM
doping. For example H-SiNW(N)+Fe is consistently ferromagnetic
semiconductor with different $E_{G,\uparrow}$ and
$E_{G,\downarrow}$. H-SiNW(N)+Mn(Cr) can be either ferromagnetic
metal or HM depending on N.

To see whether spin-dependent GGA properly represents localized
$d$-electrons and hence possible on-site repulsive Coulomb
interaction destroys the HM, we also carried out LDA+U
calculations\cite{ldau}. We found that insulating and metallic
bands of opposite spins coexist up to high values of repulsive
energy ($U=4$) for N=25. For N=57, HM persists until U$\sim$1.
Clearly, HM character of TM doped H-SiNW revealed in
Fig.\ref{fig:bands} is robust and unique behavior.

\begin{figure}
\includegraphics[scale=0.45]{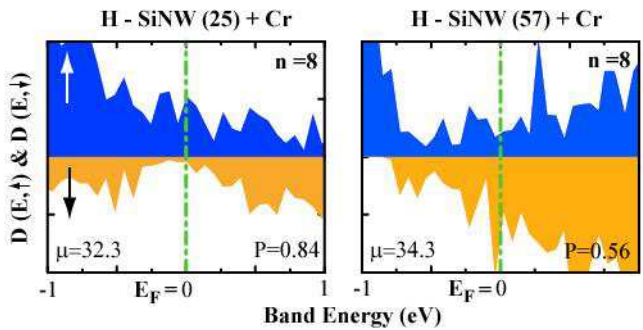}
\caption{(Color online) $D(E,\downarrow)$, density of minority (light) and
$D(E,\uparrow)$, majority (dark) spin states. (a) H-SiNW(25)+Cr, $n=8$; (b)
H-SiNW(25)+Cr, $n=16$. $P$ and $\mu$ indicate
spin-polarization and net magnetic moment (in Bohr
magnetons per primitive unit cell), respectively.}
\label{fig:dos}
\end{figure}

Finally, we note that HM state predicted in TM-doped H-SiNWs
occurs in perfect structures; complete spin-polarization may
deviate slightly from \emph{P}=100\% due to the finite extent of
devices. Even if the exact HM character corresponding to $n=1$ is
disturbed for $n>1$, the possibility that some H-SiNWs having high
spin-polarization at $E_F$ at high TM coverage can be relevant for
spintronic applications. We therefore investigated electronic and
magnetic structure of the above TM-doped H-SiNWs at $n>1$ as
described in Fig. \ref{fig:structure}(d). Figure \ref{fig:dos}
presents the calculated density of minority and majority spin
states of Cr covered H-SiNWs.

It is found that H-SiNW(21) covered by Co is non-magnetic for both
coverage of $n=4$ and 12. H-SiNW(25) is, however,
ferromagnetic for different level of Cr coverage and has high net
magnetic moment. For example, $n=8$ can be achieved by two
different geometries; both geometries are ferromagnetic with
$\mu$=19.6 and 32.3 $\mu_{B}$ and are metallic for both spin
directions. Interestingly, while $P$ is negligible for the former
geometry, the latter one has $P=0.84$ and hence is suitable for spintronic
applications (See Fig. \ref{fig:dos}). Similarly, Cr covered H-SiNW(57)
with $n=8$ and 16 are both ferromagnetic with $\mu$= 34.3
($P=$56) and $\mu$=54.5 $\mu_{B}$ ($P=$0.33), respectively.
The latter nanostructure having magnetic moment as high
as 54.5 $\mu_{B}$ can be a potential nanomagnet. Clearly, not only total
magnetic moment, but also the spin polarization at $E_{F}$ of TM
covered H-SiNMs exhibits interesting variations depending on $n$, N
and the type of TM.

In conclusion, hydrogen passivated SiNWs can exhibit half-metallic
state when doped with certain TM atoms. Resulting electronic and magnetic
properties depend on the type of dopant TM atom, as well as
on the diameter of the nanowire. As a result of TM-$3d$
and Si-$3p$ hybridization two new bands of one type of spin
direction are located in the band gap, while the bands of other
spin-direction remain to be semiconducting. Electronic properties
of these nanowires depend on
the type of dopant TM atoms, as well as on diameter of the H-SiNW.
When covered with more TM atoms, perfect half-metallic state of
H-SiNW is disturbed, but for certain cases, the spin polarization
at $E_{F}$ continues to be high. High magnetic moment obtained at high
TM coverage is another remarkable result which may lead to the
fabrication of nanomagnets for various applications. Briefly, functionalizing
silicon nanowires with TM atoms presents us a wide range of interesting
properties, such as half-metals, 1D ferromagnetic semiconductors or metals
and nanomagnets. We believe that our findings hold promise for the
use of silicon -a unique material of microelectronics- in nanospintronics
including magnetoresistance, spin-valve and non-volatile memories.


\begin{thebibliography}{99}
\bibitem{magnet}
Spin-polarized calculations have been carried by relaxing the
magnetic moment and by starting with different initial $\mu$ values.
Whether antiferromagnetic ground state exists in H-SiNW(N)+TM's
has been explored by using supercell including double primitive
cells.
\bibitem{binding}
Binding energy corresponding to $n$=1 is calculated by the
following expression,
$E_{B}=E_{T}[H-SiNW(N)]+E_{T}[TM]-E_{T}[H-SiNW(N)+TM]$ in terms of
the total energy of optimized H-SiNW(N) and TM-doped H-SiNW(N)
(i.e. H-SiNW(N)+TM) and the total energy of the string of TM atoms
having the same lattice parameter $c_{o}$ of H-SiNW(N)+TM, all
calculated in the same superlattice. Hence $E_B$ can be taken as
the binding energy of single isoalted TM atom, since the coupling
amaong adsorbed TM atoms has been excluded. To calculate
$E^{\prime}_{B}$, $E_{T}(TM)$ is taken as the total energy of bulk
TM crystal per atom. For $n$>1,
$E_T$(TM) is taken as the free TM atom energy, and hence
$\overline{E}_{B}$ includes the coupling between TM atoms. For
this reason $\overline{E}_{B}^{\prime}
> 0$ for H-SiNW(21)+Co at $n$=12
\end{thebibliography}
\end{document}